\documentclass[runningheads, a4paper]{llncs}

\usepackage[T1]{fontenc}

\usepackage{hyperref}

\usepackage{color}

\usepackage{graphicx}

\usepackage{listings}

\begin{document}
\title{Parallel Writing of Nested Data in Columnar Formats}
\author{
Jonas Hahnfeld\inst{1,2}
\and
Jakob Blomer\inst{1}
\and
Thorsten Kollegger\inst{2}
}
\authorrunning{J. Hahnfeld et al.}

\institute{
CERN, Geneva, Switzerland \\
\email{\{jonas.hahnfeld,jakob.blomer\}@cern.ch}
\and
Goethe University Frankfurt, Institute of Computer Science, Frankfurt, Germany \\
\email{kollegger@em.uni-frankfurt.de}
}
\maketitle

\begin{abstract}

High Energy Physics (HEP) experiments, for example at the Large Hadron Collider (LHC) at CERN, store data at exabyte scale in sets of files.
They use a binary columnar data format by the ROOT framework, that also transparently compresses the data.
In this format, cells are not necessarily atomic but they may contain nested collections of variable size.
The fact that row and block sizes are not known upfront makes it challenging to implement efficient parallel writing.
In particular, the data cannot be organized in a regular grid where it is possible to precompute indices and offsets for independent writing.
In this paper, we propose a scalable approach to efficient multithreaded writing of nested data in columnar format into a single file.
Our approach removes the bottleneck of a single writer while staying fully compatible with the compressed, columnar, variably row-sized data representation.
We discuss our design choices and the implementation of scalable parallel writing for ROOT's RNTuple format.
An evaluation of our approach shows perfect scalability only limited by storage bandwidth for a synthetic benchmark.
Finally we evaluate the benefits for a real-world application of dataset skimming.

\end{abstract}

\section{Introduction}
\label{sec:intro}

High Energy Physics (HEP) experiments are one of the largest scientific data producers.
One example are the experiments at the Large Hadron Collider (LHC) at CERN near Geneva:
Since the start of operation in 2010, they accumulated a total data volume of more than 2~exabytes.
This data is constantly processed and analyzed by scientists to gain new insights into the fundamental building blocks of our universe.
These analyses oftentimes only read a sparse fraction of the data.
To optimize for this use case, the data is stored in a binary columnar format implemented in the ROOT framework~\cite{ROOT}.

\begin{figure}
\centering
\begin{minipage}{0.5\textwidth}
\begin{lstlisting}[language=c++]
struct Event {
  int fId;
  std::vector<Track> fTracks;
};
\end{lstlisting}
\end{minipage}
\quad
\begin{minipage}{0.4\textwidth}
\begin{lstlisting}[language=c++]
struct Track {
  float fEnergy;
  std::vector<int> fIds;
};
\end{lstlisting}
\end{minipage}
\caption{Simplified example of nested data structures.
Real-world HEP data models often have thousands of fields.}
\label{fig:nested-data}
\end{figure}

In the coming years, the data rate is expected to increase further, for example during operation of the High-Luminosity LHC (HL-LHC).
In response, the HEP community is developing RNTuple, an evolution of the currently used TTree columnar format~\cite{BlomerEvolution}.
It is designed to make efficient use of modern hardware, but currently lacks support for highly scalable parallel writing.
This is challenging to implement efficiently because columnar formats for HEP have to handle \emph{nested data}:
Figure~\ref{fig:nested-data} shows a simplified data structure of an event consisting of many particle tracks that themselves reference other indices.
Together with transparent compression, this makes it impossible to determine row sizes upfront and to organize the data in a regular grid.
To avoid these difficulties, one commonly used solution in HEP is to independently write into separate files that are later merged.
However, this still causes overheads and efficient parallel writing would be preferable, under the assumption that rows are independent and can be reordered.

In this contribution, we present a general concept for parallel writing of nested data in columnar formats.
While developed with HEP data in mind, we believe the approach can also be applied to other columnar formats.
We discuss a concrete implementation using cooperating threads in RNTuple and evaluate it in two ways:
First we show that our implementation scales perfectly up to the storage bandwidth limit for writing randomly generated data.
Finally, we evaluate the benefits of parallel writing for a real-world application, dataset skimming, and compare our implementation to other approaches.

The remainder of the paper is structured as follows:
In Section~\ref{sec:related}, we discuss related work of other columnar data formats and existing parallel writing facilities in general, before giving an overview of RNTuple in Section~\ref{sec:rntuple}.
We present our main contribution for parallel writing of columnar data formats in Section~\ref{sec:concepts}.
In Section~\ref{sec:implementation}, we describe the implementation of parallel writing in RNTuple before evaluating its performance in Section~\ref{sec:evaluation}.
Finally, we summarize our conclusions in Section~\ref{sec:conclusions} and outline future work.

\section{Related Work}
\label{sec:related}

TTree is the current columnar data format in ROOT available for more than two decades~\cite{ROOT}.
It is used in production by HEP experiments, for example at the LHC at CERN, and stores more than 2\,EB of data.
TTree data can be written in parallel with a system called \texttt{TBufferMerger} that merges in memory the files from concurrent data producer threads~\cite{ROOTParallelIO,TBufferMerger}.
Originally \texttt{TBufferMerger} used a dedicated output thread that acted as synchronization point for all data buffers~\cite{ROOTParallelIO}.
Since then, the design has been refined to merge data directly from the worker threads~\cite{TBufferMerger}.
As a preparation to this work, we have entirely removed the data queue from \texttt{TBufferMerger}.
Instead, threads wait until they are allowed to merge, which avoids problems due to unbounded queue growth.

RNTuple is an evolution of the TTree format and currently under development~\cite{BlomerEvolution}.
It is designed as a successor to TTree that will be able to keep up with modern hardware trends and increased data rates.
As one example, RNTuple can use object store backends, for example Intel DAOS~\cite{LopezGomezObjectStores}.
At the moment, \mbox{RNTuple} is in the adoption phase by experiment frameworks.
Up to now, \mbox{RNTuple} only supports sequential writing.

Columnar data formats are also used outside of HEP and its advantages for nested data are described for the Dremel query system~\cite{Dremel}.
An open-source implementation of a columnar storage format is Apache Parquet, developed in the Hadoop ecosystem~\cite{ApacheParquet}.
It supports complex nested data structures and transparent compression with various compression algorithms.
Parquet was previously evaluated for HEP analysis needs and RNTuple was shown to outperform it for fast storage media~\cite{LopezGomezPerformance}.
To the best of our knowledge, it is not possible to write Parquet data in parallel.

NetCDF is a format for scientific data supporting fixed-size arrays and record variables with a single unlimited dimension~\cite{NetCDF}.
This makes it possible to implement efficient random access by computing offsets into the record variables.
A parallel interface exists in the form of PnetCDF, which builds on top of MPI-IO~\cite{PnetCDF}.
However, all record variables are stored interleaved and assumed to grow together, which is not suitable for nested data.

HDF5 features a rich data model with extensible datasets, nested datatypes, as well as variable length arrays~\cite{HDF5}.
It supports parallel writing via MPI-IO, but at the time of writing not together with variable length arrays.
For extensible datasets, it has to be noted that the method \texttt{H5Dset\_extent} is a collective operation.
This means it is not suitable for independent parallel writes of nested data in columnar formats.

\section{RNTuple Overview}
\label{sec:rntuple}

In this section, we give a short overview of the RNTuple format and terminology~\cite{BlomerEvolution} as needed for this work.
For serializing acyclic C++ data structures, \mbox{RNTuple} decomposes them recursively into \emph{fields}.
Variable-length collections are handled by transformation into two fields:
An offset field points into a second field holding the element data, which may be recursively decomposed further.
Table~\ref{tab:nested-data} shows an example columnar representation for the nested data structures presented in Figure~\ref{fig:nested-data}.
\texttt{fTracks} and \texttt{fTracks.\_0.fIds} point into their \texttt{\_0} subfields and past the last element included in the \texttt{vector}s.
The leafs of this field tree are mapped to \emph{columns} of primitive, fixed-size types.
Each column is partitioned in \emph{pages}.
Elements of the columns are written consecutively into pages.
Pages are also the units of compression.
At the moment, ROOT supports the compression algorithms DEFLATE~\cite{rfc1951}, LZMA~\cite{lzma}, LZ4~\cite{lz4}, and Zstandard~\cite{rfc8878}.

\begin{table}
\centering
\caption{Example of columnar representation for the nested data structure shown in Figure~\ref{fig:nested-data}. Gaps and lines are inserted for clarity only.}
\label{tab:nested-data}
\begin{tabular}{|c|c|c|c|c|}
\hline
\texttt{fId} & \texttt{fTracks} & \texttt{fTracks.\_0.fEnergy} & \texttt{fTracks.\_0.fIds} & \texttt{fTracks.\_0.fIds.\_0} \\
\hline
6873 & 2 & 25.4f & 2 & 42 \\
 & & & & 27 \\
 & & 32.8f & 3 & 16 \\\hline
6874 & 3 & 14.7f & 5 & 21 \\
 & & & & 8 \\\hline
\vdots & \vdots & \vdots & \vdots & \vdots \\
\end{tabular}
\end{table}

A row or an \emph{entry} (also called \emph{event} for HEP data) spans all corresponding field and column elements of the potentially nested data.
For example, Table~\ref{tab:nested-data} shows two entries separated by lines for clarity.
In the current implementation, all pages of a consecutive range of entries form a \emph{cluster}
\footnote{
The RNTuple specification allows for sharded clusters of subsets of columns, but this is currently not used.
}.
The order between pages of a cluster is established via a \emph{page list}.
It stores the number of elements in each page and where it is located, for example at a given byte offset.
The page list also contains the element offset for each column in a cluster.
Together with the sum of all elements, this defines the \emph{column range} of the column elements in a cluster.

RNTuple data is meant to be embedded into a container format, for example into ROOT files.
Also known as a \texttt{TFile}, a ROOT file is a self-describing binary format consisting of a linked list of \emph{keys}.
Every key reserves a byte buffer that, for RNTuple, is referenced as a byte offset.
It is possible to implement embeddings into other container formats and also object stores~\cite{LopezGomezObjectStores}.

\section{Concepts for Parallel Writing of Columnar Data}
\label{sec:concepts}

Columnar formats with support for nested data inevitably have variable row sizes.
Moreover, it is generally not possible to predict the compression ratio upfront in case of transparent compression.
As a result, the data cannot be organized in a regular grid where chunks can be processed independently.
Instead, we propose a more dynamic approach for parallel writing with cooperating threads.

To explain our approach, we first introduce the concept of a \emph{unit of writing}:
It is a consecutive block of the data in the serialized and compressed on-disk representation.
For parallel writing, a unit of writing defines the granularity of synchronization and thereby strongly influences scalability.
For RNTuple, possible units of writing can be a single page or all pages of a cluster.

\subsection{Serialization and Compression}
\label{sec:serialization-compression}

Data serialization, and especially data compression, are time-consuming operations with a throughput of less than 1\,GB/s~\cite{zstd}.
One possible approach to parallelization is to distribute the serialization and compression of one unit of writing to multiple threads.
As we will show later, it is however more efficient and scalable if each thread works on their own unit of writing.

Parallel writing is possible under the assumption that a unit of writing is \emph{relocatable}.
With this, we mean that the serialized and compressed unit can be moved without requiring changes in its contents.
In that case, serialization and compression can happen without synchronization.
This allows parallel writing to independently prepare multiple units of writing and store them at any location.

\subsection{Writing into Container Format}
\label{sec:writing-container}

After serialization and compression, the unit's final size is known.
Based on this information, parallel writing can reserve an appropriate buffer in the container format.
In general, this step requires synchronization to allocate the necessary resources.
For example, in the case of writing into files, this is needed to ensure a linear file layout.

Synchronization requirements for the actual writing depend on the destination.
For local files, multiple threads can write their unit in parallel into previously reserved offsets.
The operating system aggregates and schedules the writes depending on hardware capabilities.
For distributed systems, such as parallel filesystems and object stores, the achievable parallelism depends on the mapping to backend servers.
In either case, it is also possible to serialize and compress in parallel but to linearize the write operations with a critical section.
This simplifies the implementation requirements and can already be sufficient to achieve good scalability.

\subsection{Updating Format Metadata}
\label{sec:format-data}

Once the unit of writing is stored, the parallel writer needs to update format-specific metadata information.
To ensure compatibility with reader implementations, this should happen as if the data was written sequentially and generally requires synchronization.
For \mbox{RNTuple}, this step includes updating the column ranges and adding locators to the page list.

\section{Implementation of Parallel Writing in RNTuple}
\label{sec:implementation}

Based on the concepts described above, we implement multithreaded writing for ROOT's RNTuple format.
As unit of writing, we choose clusters that are prepared in parallel by multiple threads.
Clusters are relocatable because the offset columns are relative to the current cluster.
The compressed pages are buffered in memory until the entire cluster is ready to be committed.
At that point, the thread takes a lock, writes all pages to storage, and updates the metadata, namely the page list and column ranges.

It is also possible to choose a single page as unit of writing for RNTuple.
In that case, compressed pages are directly written to storage and the thread only remembers the byte location for later updating of the metadata.
While this reduces memory consumption, it requires locking per page and turns out to scale badly due to lock contention.
We will discuss a detailed assessment in Section~\ref{sec:scalability}.

During the implementation, we find two performance optimizations that can improve scalability:
First, for some filesystems it is beneficial to reserve the space taken by all compressed pages in a cluster before starting to write.
On POSIX systems this can be done with the \texttt{posix\_fallocate}
\footnote{
On Linux systems, the platform-specific \texttt{fallocate} may be more appropriate because it avoids emulation in case the underlying filesystem does not support the system call.
}
function which allows the filesystem to pre-allocate blocks of sufficient size.
We will present performance numbers for this optimization in Section~\ref{sec:scalability}.
As a follow-up to this optimization, we note that it is not necessary to write the pages to storage while holding the lock.
Instead, it is sufficient to just reserve the space and fill the page locations in the metadata.
Afterwards the data can be written truly in parallel outside the critical section.
We implemented both optimizations for test purposes.
Where we see a significant improvement due to pre-allocating space, the evaluation points it out.
Otherwise, the evaluation leaves out both potential optimizations.

\section{Evaluation of Parallel RNTuple Writing}
\label{sec:evaluation}

In this section, we evaluate our implementation of parallel RNTuple writing.
We first discuss results of a synthetic benchmark on different storage media in Section~\ref{sec:scalability}.
Afterwards we show an application benchmark of dataset skimming in Section~\ref{sec:skim-agc}.
The corresponding source code is openly available on GitHub~\cite{rntuple-apps}.

All tests are executed on a dedicated benchmarking server with a single AMD EPYC 7702P processor running at a fixed frequency of 2.0\,GHz.
It has 64~physical cores with simultaneous multithreading (SMT) for a total of 128~threads.
The server runs AlmaLinux~9.3 with the stock kernel 5.14.0-362.18.1.
Deviating from the default configuration, we turn on \texttt{io\_uring}
\footnote{
\texttt{sysctl kernel.io\_uring\_disabled=0}
}
which RNTuple can use for better performance when reading data.
We compile a development version of ROOT in a \texttt{Release} configuration using GCC~11.4.1.
Unless stated otherwise, we compress the RNTuple data with the default zstd algorithm.

\subsection{Scalability on Different Storage Media}
\label{sec:scalability}

To assess the scalability of parallel RNTuple writing, we first test its weak scaling behavior with up to 128~threads for a synthetic benchmark.
Each thread writes a fixed number of entries with two fields: an ``event ID'' and a vector of particles.
The vector is sized following a Poisson distribution with mean $\mu = 5$ and filled with \texttt{float}s drawn from a uniform distribution $[0, 100)$.
Increasing the amount of data with the number of threads is based on the assumption that HEP experiments have many events to process.
Without parallel writing, the events would be distributed to multiple processes and written into multiple smaller files.
To reduce the variance of the results, we execute each configuration five times and report the harmonic mean of the bandwidth.
We verify that at each point the margin of error of the 95\,\% confidence interval is smaller than 5\,\% of the mean.

\subsubsection{Writing to \texttt{/dev/null}}

\begin{figure}
\centering
\includegraphics{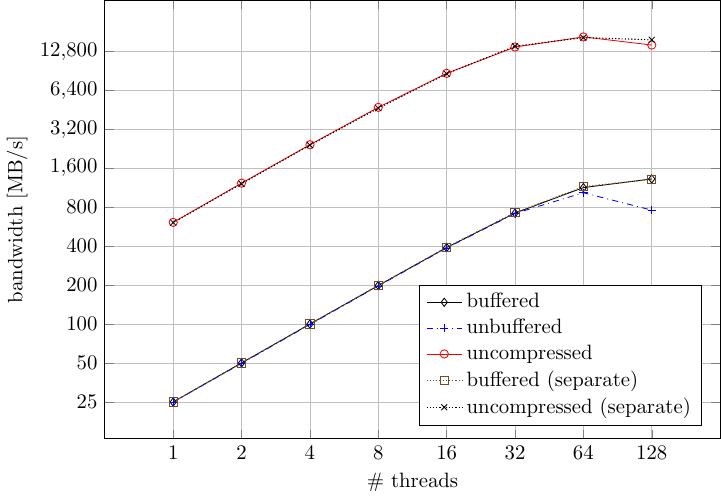}
\caption{Bandwidth measured with the synthetic benchmark writing to \texttt{/dev/null}.
Every thread writes 20~million entries.}
\label{fig:scalability-null}
\end{figure}

To test the scenario of infinitely fast storage, we first write the RNTuple data into \texttt{/dev/null}.
In that way, the benchmark involves all parts of the userspace software stack including system calls into the kernel.
Figure~\ref{fig:scalability-null} shows the computed bandwidth from writing 20~million entries per thread.
For the bottom lines, 20~million entries amount to 337\,MB of compressed data per thread and the bandwidth is limited by compression speed.
In the default ``buffered'' configuration, the unit of writing is an entire cluster and compressed pages are buffered in memory as described in Section~\ref{sec:implementation}.
It can be seen that parallel writing scales from 25\,MB/s for one thread to 1135\,MB/s~(45.4x) with 64~threads and 1324\,MB/s~(53x) with 128~threads.
To investigate the non-linear scaling at higher thread counts, we test a variant where each thread creates a separate sequential writer.
As shown with the brown squares in the plot, the bandwidths are identical to parallel writing.

In contrast, the ``unbuffered'' configuration writes compressed pages directly to storage, which requires taking a lock per page.
It can be seen that this works for lower thread counts and gives near-identical bandwidths up to 32~threads.
However, at 64~threads and especially with SMT, locking per page results in lower bandwidths because of lock contention.
We confirm this hypothesis by counting the number of \texttt{futex} system calls using \texttt{strace}:
For 64~threads, the ``buffered'' configuration makes around 300~system calls while the ``unbuffered'' has more than 27,000.

For comparison, the red line shows the bandwidth of writing the same 720\,MB of uncompressed data per thread.
It scales from 612\,MB/s with one thread to 16,553\,MB/s (27.1x) with 64~threads and slightly decreases to 14,330\,MB/s with 128~threads.
As in the buffered case, this behavior is similar to the bandwidths measured with a separate writer per thread.

\subsubsection{Writing to SSD}

\begin{figure}
\centering
\includegraphics{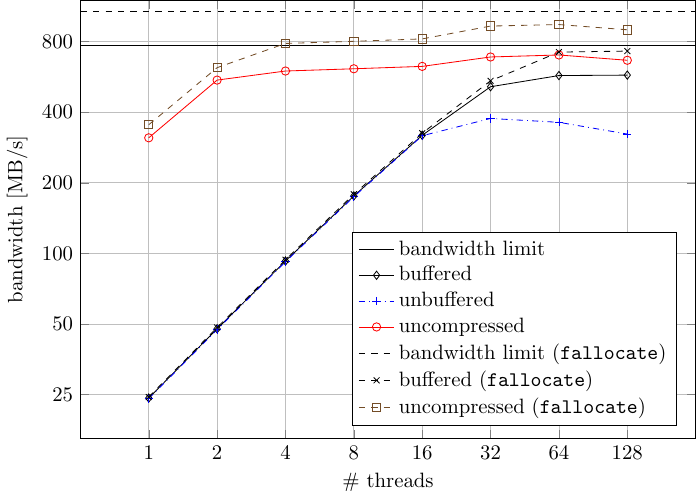}
\caption{Bandwidth measured with the synthetic benchmark writing on a server SSD.
Every thread writes 20~million entries.}
\label{fig:scalability-ssd}
\end{figure}

In a second setup, we write the synthetic data on a Samsung PM1733 NVMe SSD formatted with \texttt{ext4}.
Figure~\ref{fig:scalability-ssd} presents the results.
As before, every thread writes 20~million entries.
First we measure two different limits with the Flexible I/O Tester (\texttt{fio})~\cite{fio} and draw them as horizontal lines:
The lower solid one at 771\,MB/s uses a blocksize of 64\,KiB and extends the file size while writing.
When instead preallocating the file with \texttt{fallocate}, we reach the upper bandwidth of 1075\,MB/s drawn as a dashed line.

By default, RNTuple targets an uncompressed page size of 64\,KiB.
This makes it possible to directly compare the uncompressed configurations with the bandwidth limit measured by \texttt{fio}.
We find that our parallel writing implementation achieves close to 90\,\% of that limit at 64~threads:
Without further optimization, the bandwidth peaks at 702\,MB/s (91\,\% of the limit) in the red curve.
If instead preallocating the size of all pages in a cluster before writing, the implementation reaches 947\,MB/s (88\,\%) as shown with the brown curve.

With zstd compression enabled, the bandwidth first increases linearly comparable to Figure~\ref{fig:scalability-null}.
At 16~threads, the curves start to flatten out and eventually reach a plateau of 576\,MB/s and 729\,MB/s, respectively.
As before, the ``unbuffered'' configuration achieves worse bandwidths at higher thread counts due to lock contention.

\subsubsection{Writing to HDD}

\begin{figure}
\centering
\includegraphics{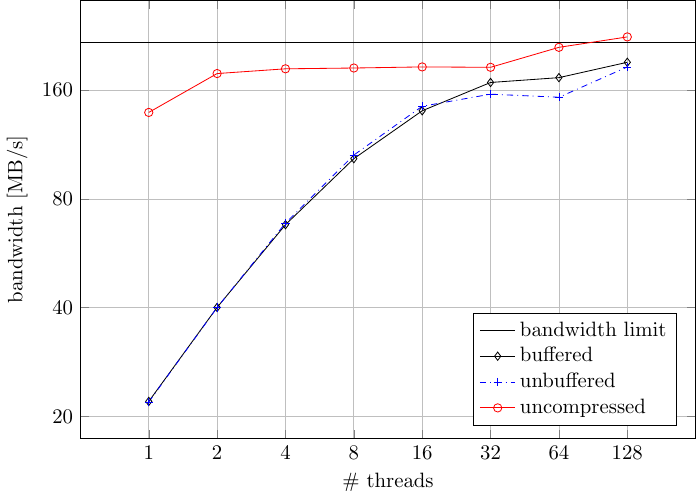}
\caption{Bandwidth measured with the synthetic benchmark writing on a server HDD.
Every thread writes 10~million entries.}
\label{fig:scalability-hdd}
\end{figure}

Finally, we test parallel writing to a slower spinning disk, a Toshiba MG07ACA also formatted with \texttt{ext4}.
Figure~\ref{fig:scalability-hdd} presents the results.
Because of lower expected bandwidths, we decrease the number of entries to 10~million per thread.
Again we first measure a bandwidth limit of 217\,MB/s using \texttt{fio}.
It can be seen that, without compression, parallel writing reaches a plateau of around 180\,MB/s already with two threads.
The bandwidth increases to 225\,MB/s at 128~threads and thereby slightly overshoots the limit.
With compression enabled, the ``buffered'' configuration reaches 191\,MB/s at 128~threads.
In this case, the ``unbuffered'' configuration achieves a similar bandwidth of 186\,MB/s.
We do not present numbers with \texttt{fallocate} as they are identical to the measurements without the optimization.

\subsection{Dataset Skimming of the Analysis Grand Challenge}
\label{sec:skim-agc}

The Analysis Grand Challenge (AGC)~\cite{AGC} aims to test analysis workflows at scales required for experiments at the future, upgraded LHC accelerator (HL-LHC)
that will take data in the 2030s.
The AGC both specifies a typical HEP physics analysis (a ``ttbar'' analysis) that uses Open Data as input, and it provides a reference implementation~\cite{AGCdocs}.
The input dataset is derived from 2015 Open Data of the CMS experiment and openly available in the TTree format.
After conversion to RNTuple with zstd compression, the total data volume amounts to 969\,GB across 787~files.

The dataset is partitioned in events of different \emph{physics processes} and, for some processes, \emph{variations} where one variable is scaled or \emph{varied}.
In total the AGC analysis dataset has nine partitions.
To evaluate parallel writing, our application pre-filters the data of the various input files and produces nine output files, one for each partition.
Such a data preparation task is a typical procedure in the course of a physics analysis.

To reduce the size of the dataset, we apply three strategies:
First we drop unneeded columns and only retain fields that are actually read and used by the analysis.
This is commonly known as \emph{horizontal skimming} and in our case decreases the dataset size to 56\,GB.
Second we apply \emph{vertical skimming} and filter out events (entries) based on coarse cuts:
Version 1 of the AGC analysis task requires exactly one electron or muon with transversal momentum $p_T > 25$\,GeV to be present in an event.
Additionally, only events are taken into consideration with at least four jets of $p_T > 25$\,GeV
\footnote{
A second version of the analysis task is currently worked on that increases the cut values to $p_T > 30$\,GeV and adds additional constraints.
The exact details are not of utmost relevance for this work, so we decide to stick to the initial, frozen version.
}.
Therefore, we only retain events with at least one electron and muon and at least four jets with $p_T > 20$\,GeV.
This reduces the dataset size further to 24\,GB while not changing the results of the analysis.
Finally, for events that passed the cut, we also drop entries from the nested collections of electrons, muons, and jets with $p_T < 20$\,GeV.
With all three skims combined, we end up with a dataset of nine files with a total size of 19\,GB.

\begin{figure}
\centering
\includegraphics{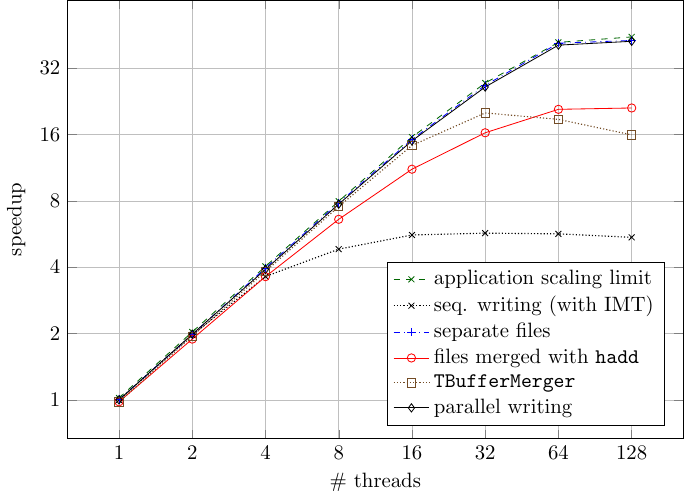}
\caption{Speedup of the AGC dataset skimming benchmark compared to a full sequential run of 2432~seconds.}
\label{fig:agc-speedup}
\end{figure}

We compare the parallel \mbox{RNTuple} writing against four different alternative ways of writing the data.
In all cases, the skimmed files are written to the same Samsung PM1733 NVMe SSD as used before, and without the \texttt{fallocate} optimization.
We execute each application configuration five times and take the arithmetic mean of the runtimes.
At each point, we verify that the margin of error of the 90\,\% confidence interval is smaller than 5\,\% of the mean.
Figure~\ref{fig:agc-speedup} shows the resulting speedup plot compared to a full sequential run.

All versions have in common that they parallelize over output partitions as the partitions are independent of each other.
We use Intel Threading Building Blocks~\cite{TBB} to express the parallelism, accessed via auxiliary classes provided by ROOT.
This allows to control the total number of threads available to the benchmark task.

As a first option, we do not parallelize over the input files, but instead turn on ROOT's \emph{implicit multithreading} (IMT).
The sequential RNTuple writing uses this facility to parallelize page compression.
It can be seen from the plot that this method scales up to four threads, but then reaches a plateau at around 430~seconds.
This corresponds to a speedup of factor 5.7x shown with the black dotted curve compared to the full sequential run of 2432~seconds.

A different approach is to skim each input file into a separate output file with fully independent writers.
As expected, this approach scales up to high thread counts and finishes in 57~seconds when using 128~threads.
However, it produces 787~files instead of one for each of the nine partitions.
It is possible to merge files using ROOT's \texttt{hadd} utility as a postprocessing step.
This requires reading back the written files and takes around 58~seconds when parallelizing over the output partitions.
When adding this time to the skimming itself, the speedup is worse as shown by the red curve compared to the blue curve in the plot.
An additional disadvantage is that it requires double the storage space while the merge is ongoing.

As discussed in Section~\ref{sec:related}, \texttt{TBufferMerger} allows for merging files in memory.
For this work, we extended this functionality to also merge RNTuple data.
This allows to parallelize over the input files, but still only produce one output file per partition.
We find that this version scales up to 32~threads with an absolute runtime of 121~seconds in the brown curve.
For higher thread counts, its runtime starts to increase again.

Finally, the black solid curve shows the implementation using the new parallel RNTuple writing.
We find that it scales as well as independent writing to separate files.
At 128~threads, it skims the dataset in 57~seconds which is a factor 42.7x faster than the full sequential run.
For writing the nine output files with a total of 19\,GB, this corresponds to a bandwidth of 330\,MB/s, which is below the limit determined with the synthetic benchmark.
As an additional test, we run the variant with separate output files per thread but write into \texttt{/dev/null}.
This allows to determine the scalability limit of the application indicated by the green curve in the plot, with the best runtime of 55~seconds using 128~threads.
We therefore conclude that the application is not bound by the scalability of the parallel RNTuple writing.

\section{Conclusions and Future Work}
\label{sec:conclusions}

In this paper, we presented a concept for parallel writing of nested data in columnar formats.
We discussed implementation choices and performance optimizations for the RNTuple format.
Finally, we evaluated our implementation with a synthetic benchmark of writing randomly generated data and an application of dataset skimming.

With the synthetic benchmark, we showed that parallel writing of RNTuple data scales up to the storage bandwidth limit.
In order to increase that limit, we plan to investigate the usage of Direct I/O, which results in better performance using \texttt{fio}.
Unfortunately first tests with \mbox{RNTuple} showed that Direct I/O cannot be simply turned on because of its strict alignment requirements.
More work will be needed to write the buffered pages of a cluster in appropriate chunks and tune the implementation.

Furthermore, we want to investigate if parallel RNTuple writing can be extended to process-level parallelism on the same node.
This would allow more flexibility in mixing process-level and thread-level parallelism, which some HEP experiments exploit.
For this, we plan to use the second optimization described in Section~\ref{sec:implementation} to decouple writing from the critical section.
In that way, it will be possible to reduce synchronization overhead by only transferring metadata, but keeping the column data local to each process.
If successful, a further step would be distributed parallelism with MPI and writing to cluster filesystems.
Finally, we want to test our implementation with more applications
and address potential integration issues when combining parallel writing with other parallel components of HEP experiment software.

\begin{credits}
\subsubsection{\ackname}
This work has been sponsored by the Wolfgang Gentner Programme of the German Federal Ministry of Education and Research (grant\,13E18CHA).

\subsubsection{\discintname}
The authors have no competing interests to declare that are relevant to the content of this article.
\end{credits}

\bibliographystyle{splncs04}
\bibliography{paper}
\end{document}